%% file: monniaux_bodin_aplas2011_article.tex
\documentclass[a4paper]{article}
\usepackage{amsmath,amssymb,amsfonts}
\usepackage{listings}
\usepackage[numbers]{dmnatbib}

\usepackage{tikz}
\usetikzlibrary{trees,arrows,automata,shapes}
\usepackage{hyperref}

\title{Modular Abstractions of Reactive Nodes using Disjunctive Invariants%
\thanks{This work was partially supported by ANR project ``\href{http://asopt.inrialpes.fr/}{ASOPT}''.}}

\author{David Monniaux\thanks{\href{http://www.cnrs.fr/}{CNRS} / \href{http://www-verimag.imag.fr/}{VERIMAG}} \and Martin Bodin\thanks{\href{http://www.ens-lyon.eu/}{\'Ecole normale sup\'erieure de Lyon}; on internship at CNRS / VERIMAG.}}

\newcommand{\true}{\textsf{true}}
\newcommand{\false}{\textsf{false}}

\newcommand{\bbZ}{\mathbb{Z}}
\newcommand{\defn}{\stackrel{\triangle}{=}}
\newcommand{\calT}{\mathcal{T}}
\newcommand{\frakT}{\mathfrak{T}}
\newcommand{\calI}{\mathcal{I}}
\newcommand{\calO}{\mathcal{O}}

\usepackage{lustrelisting}
\newcommand{\soft}[1]{\textsc{#1}}

\begin{document}
\input{monniaux_bodin_aplas2011}

\phantomsection\addcontentsline{toc}{section}{References}
\bibliographystyle{plainnat}
\bibliography{monniaux_bodin_aplas2011}
\end{document}

%% file: monniaux_bodin_aplas2011.tex
\maketitle

\begin{abstract}
We wish to abstract nodes in a reactive programming language, such as Lustre, into nodes with a simpler control structure, with a bound on the number of control states. In order to do so, we compute disjunctive invariants in predicate abstraction, with a bounded number of disjuncts, then we abstract the node, each disjunct representing an abstract state. The computation of the disjunctive invariant is performed by a form of quantifier elimination expressed using SMT-solving.

The same method can also be used to obtain disjunctive loop invariants.
\end{abstract}

\section{Introduction}
Our goal is to be able to compute sound abstractions of reactive nodes, with tunable precision. A reactive node in a language such as
\soft{Lustre},%
\footnote{\soft{Lustre} is a synchronous programming language, which gets compiled into~C.~\cite{LUSTRE_POPL87}.} or
\soft{Scade},%
\footnote{\soft{Scade} is a graphical synchronous programming language derived from \soft{Lustre}. It is available from \href{http://www.esterel-technologies.com/}{Esterel Technologies}. It has been used, for instance, for implementing parts of the Airbus A380 fly-by-wire systems.}
\soft{Sao},%
\footnote{\soft{Sao} is an earlier industrial graphical synchronous programming language. It has been used, for instance, for implementing parts of the Airbus A340 fly-by-wire systems.}
or even \soft{Simulink},%
\footnote{\soft{Simulink} is a graphical data-flow modeling tool sold as an extension to the \soft{Matlab} numerical computation package. It allows modeling a physical or electrical environment along the computerized control system. A code generator tool can then provide executable code for the control system for a variety of targets, including generic~C. \soft{Simulink} is available from \href{http://www.mathworks.com/}{The Mathworks}.}
has input streams, output streams, and an (optional) internal state: at each clock cycle, the value on each output is a function of the values on the inputs and the state; and so is the next value of the state.

If the state consists in a finite vector of Booleans, or other finite values, then the node is a finite automaton, with transitions guarded according to the current values of the inputs, and for each state a relation between the current values of the inputs and the current values of the outputs. This is often referred to as the \emph{control structure} of the reactive program. The problem with that representation, which exposes the full internal state, is that the number of states grows exponentially with the number of state variables, making it unwieldy for analysis. The problem is even more severe if the control conditions are not directly exposed as Boolean state variables, but as predicates over, say, integer or real variables (see example in Sec.~\ref{part:construct_automaton}).

The main contribution of this article is a method for constructing a more abstract automaton, with a bounded number of states ($\leq n$), whose behaviors still over-approximate the behaviors of the node. In order to do so:
\begin{enumerate}
\item We compute an over-approximation of the set of reachable states of the node, in an unspecified context, as a union of at most $n$ ``abstract states'', each defined by a conjunction of constraints (these abstract states need not be disjoint).\label{item:analysis}
\item We compute the most precise transition relation between these abstract states.
\end{enumerate}

This automatic abstraction maps a reactive node into another, more abstract (and, in general, nondeterministic) reactive node. This enables modular and compositional analysis: if a node is composed of several nodes, then one can replace each of these nodes by its abstraction, and then analyze the compound node.

As a secondary contribution, the analysis method at step~\ref{item:analysis} can also be used to obtain disjunctive loop invariants for imperative programs (or, more generally, invariants for arbitrary control flow graphs), given a precondition and an optional postcondition. We describe this algorithm for obtaining invariants in disjunctive normal form, but it in fact also works for other templates.

Our algorithms use \emph{satisfiability modulo theory} (SMT) solving as an essential subroutine; see e.g. \cite{Cimatti_WODES08} for an introduction.

\section{Invariants by Predicate Abstraction}
Predicate abstraction abstracts program states using the truth value of a given finite set of predicates $\{ \pi_1, \dots, \pi_m \}$: each state $\sigma$ is abstracted by a $m$-tuple of Booleans$(\pi_1(\sigma), \dots, \pi_m(\sigma))$.
The most precise abstract transition relation between such vectors of Booleans is $(B_1, \dots, B_m) \rightarrow_\pi (B'_1, \dots, B'_m)$ if and only if there exist $\sigma \models \bigwedge (\pi_i = B_i)$, $\sigma' \models \bigwedge (\pi_i = B'_i)$, and $\sigma \rightarrow \sigma'$ where $\rightarrow$ is the transition relation of the program. Then, given an abstract initial state, the set of reachable states of the abstract transition relation can be computed within finite time (in general, exponential in~$m$) by Kleene iterations (equivalently, by computing the transitive closure of $\rightarrow_\pi$).

Such an approach is, however, unworkable in general because of the exponential number of states generated, and thus all current predicate abstraction schemes use some stronger form of abstraction \cite{Graf_Saidi_CAV97}; for instance, they may simply compute a conjunction of the $\pi_i$ that holds inductively at a given program point. Conjunctive invariants are however fairly restrictive; in this article, we consider the problem of obtaining invariants as \emph{disjunctions of a fixed number of conjunctions} of the chosen predicates.

The set of reachable states of a reactive node, in an unspecified environment, is the strongest invariant of an infinite loop:
\lstset{language=Java}
\begin{lstlisting}
while (true) {
  i = inputs();
  o = outputs(state, i);
  state = next_state(state, i);
}
\end{lstlisting}

We shall therefore investigate the problem of automatically finding disjunctive inductive loop invariants (or, more generally, invariants for predicate abstraction following a fixed template), using predicate abstraction, given a precondition and an optional postcondition. These invariants shall be minimal with respect to the inclusion ordering: there shall be no stronger inductive invariant definable by the same template.

\subsection{Solution of a Universally Quantified Formula}
Let us assume a finite set $\Pi = \{ \pi_1, \dots, \pi_m \}$ of predicates over the state space of the variables of the program. Let $n \geq 1$ be an integer. We are looking for invariants of the form $C_1 \lor \dots \lor C_n$ where the $C_i$ are conjunctions of predicates from~$\Pi$ (most of our techniques are not specific to this template form, see Sec.~\ref{sec:other_forms} for extensions).

Any such invariant can be obtained by instantiating the Booleans $b_{i,j}$ in the following template:
\begin{equation}
\calT \defn \bigvee_i \underbrace{\bigwedge b_{i,j} \Rightarrow \pi_j}_{C_i}
\end{equation}
Setting $b_{i,j}$ to \true (respectively, \false) in that template means that predicate $\pi_j$ appears (respectively, does not appear) in the $i$-th disjunct $C_i$. For instance, if $\Pi = \{ x > 0, x < 1, y > 0 \}$ and $n=2$, then $b_{1,1}=\true$, $b_{1,2}=\true$, $b_{1,3}=\false$, $b_{2,1}=\false$, $b_{2,2}=\false$, $b_{2,3}=\true$ correspond to $(x > 0 \land x < 1) \lor y > 0$.

The problem of finding an invariant reduces to finding suitable values for these Booleans.
There is therefore a search space for invariant candidates of \emph{a priori} size $2^{mn}$. We impose that the invariant $I$ obtained be \emph{minimal} within that search space with respect to the inclusion ordering; that is, there is no $I'$ expressive using the template such that~$I' \subsetneq I$.

Our algorithm can in fact apply to any control-flow graph. For the sake of simplicity, we shall describe it on a single loop.
\medskip

In Hoare logic, the conditions for proving that a postcondition $P$ holds after a while loop whose condition is $C$, whose transition relation is $T$ and whose precondition is $S$ using loop invariant $I$ are:
\begin{itemize}
\item $I$ must contain the precondition, otherwise said $\forall \sigma ~ S \Rightarrow I$.
\item $I$ must be inductive, otherwise said $\forall \sigma, \sigma' ~ I \land C \land T \Rightarrow I'$, with $I'$ denoting $I$ where all state variables have been primed.
\item $I \land \neg C$ must imply the postcondition, otherwise said $\forall \sigma~ I \land \neg C \Rightarrow P$.
\end{itemize}

If we impose $I$ to be an invariant of the required form, that is, an instantiation $\calT[B/b]$ of $\calT$ obtained by setting the $b_{i,j}$ variables to certain values $B_{i,j}$, these conditions boil down to the values  $B_{i,j}$ of the $b_{i,j}$ variables must satisfy certain formulas universally quantified over the state $\sigma$ or on the couple of states $\sigma,\sigma'$.

We now make an additional assumption: the states $\sigma$ or $\sigma'$ comprise a fixed, \emph{finite} number of variables%
\footnote{These variables are not necessarily scalar variables. It is for instance possible to consider uninterpreted functions from the integers to the integers, which stand for a countably infinite number of integers.}
expressible in a theory $\frakT$ for which there exists a satisfiability testing algorithm, in which the predicates $\pi_1, \dots, \pi_m$ can be expressed, and which allows propositional variables. Thus, the problem boils down to finding a solution to a conjunction of universally quantified formulas of that theory such that the only free variables are the $b_{i,j}$ Booleans.

In the following sections, lowercase $\sigma$ and $\sigma'$ stand for states (thus stand for a finite number of variables in the theory $\frakT$), uppercase $\Sigma$ and $\Sigma'$ stand for values of these state variables. Similarly, lowercase $b$ stands for the matrix of propositional variables $(b_{i,j})_{1 \leq i \leq m, 1 \leq j \leq n}$, and uppercase $B$ stands for the matrix of Booleans $(B_{i,j})_{1 \leq i \leq m, 1 \leq j \leq n}$. $F[B/b]$ thus stands for the formula $F$ where the propositional values $b$ have been replaced by the corresponding Booleans in $B$, and $F[\Sigma/\sigma]$ stands for the formula $F$ where the state variable $\sigma$ has been replaced by the state value~$\Sigma$.

\subsection{Naive Algorithm for a Given Postcondition}
\label{part:with_postcond}

In this section, we shall explain how to compute an invariant suitable for proving the Hoare triple of a loop, given a precondition, a postcondition (which may be \true), a loop condition and a transition relation.

Let us first give an intuition of the algorithm. A universally quantified formula $\forall \sigma F$ with free Boolean variables $b$ can be understood as specifying a potentially infinite number of constraints $F[\Sigma/\sigma]$ over $b$, where $\Sigma$ ranges all possible values for~$\sigma$ (in this section, we will lump together $\sigma$ and $\sigma'$ as a single~$\sigma$). The idea is to ``discover'' such constraints one at a time, when they are violated. 

Let us now examine the algorithm in more detail; see Sec.~\ref{sec:step_by_step_example} for a complete algorithm run.
The $H_k$ sequence of propositional formulas over the $b$ variables will express successive refinements of the constraints during the search of a suitable assignment. Initially, we do not know anything about possible solutions, so we set $H_1 \defn \true$.

We start by taking any initial assignment $B^{(1)}$  (since any will satisfy $H_1$) and check whether $\neg F[B^{(1)}/b]$ is satisfiable, that is, whether one can find suitable values for $\sigma$. If it is not, then $B^{(1)} \models \forall \sigma~F$. If it is satisfiable, with example value $\Sigma_1$, we add $F[\Sigma_1/\sigma]$ as a constraint --- that is, we take $H_2 \defn H_1 \land F[\Sigma_1/\sigma]$; note that this constraint excludes $B^{(1)}$ and possibly other values for $b$. Now find an assignment $B^{(2)}$ satisfying $H_2$, check whether $\neg F[B^{(2)}/b]$ is satisfiable. If it is not, then $B^{(2)} \models \forall \sigma~F$. If it is satisfiable, with example value $\Sigma_2$, we take $H_3 = H_2 \land F[\Sigma_2/\sigma]$; note that $H_3$ excludes $B^{(1)}$ and $B^{(2)}$. The process continues until a suitable assignment is found or the constraints exclude all assignments. Note that one Boolean assignment at least is excluded at each iteration, and that the number of Boolean assignments is finite (exponential in the number of propositional variables in~$b$).

More formally: recall that we have reduced our problem of finding an invariant to finding Boolean values $B_{i,j}$ such that $(B_{i,j})_{1 \leq i \leq m, 1 \leq j \leq n} \models \forall \sigma~F$ for a certain quantifier-free formula $F$ whose free variables are $(b_{i,j})_{1 \leq i \leq m, 1 \leq j \leq n}$. Let us now assume we have a SMT-solver for theory $\frakT$, a function $\textit{SMT}(G)$ which given a formula $G$ answers $\textit{sat}(M)$ when $G$ is satisfiable, where $M$ is a model, that is, a suitable instantiation of the free variables in $G$, or $\textit{unsat}$ otherwise. We shall also assume a SAT-solver $\textit{SAT}$ with similar notations, for purely propositional formulas. We run the following algorithm, expressed in pseudo-ML:
\bigskip

\noindent
H := \true\\
\textbf{loop}\\
\hspace*{1em} \textbf{match} \textit{SAT}(H) \textbf{with}\\
\hspace*{1em} $|$ \textit{unsat} $\rightarrow$ \textbf{return} ``no solution''\\
\hspace*{1em} $|$ $\textit{sat}((B_{i,j})_{1 \leq i \leq m, 1 \leq j \leq n})$ $\rightarrow$\\
\hspace*{2em} \textbf{match} $\textit{SMT}(\neg F[B/b])$ \textbf{with}\\
\hspace*{2em} $|$ \textit{unsat} $\rightarrow$ \textbf{return} ``solution~$B$''\\
\hspace*{2em} $|$ $\textit{sat}(\Sigma)$ $\rightarrow$ $H := H \land F[\Sigma/\sigma]$.
\bigskip

This algorithm always terminates, since the main loop iterates over a finite set of size $2^{|b|}$ where $|b|=mn$ is the size of the matrix $b$ of propositional variables: the number of models of the propositional formula $H$ decreases by at least one at each iteration, since model $B$ is excluded by the $F[\Sigma/\sigma]$ condition. The loop invariant is $\forall \sigma~F \implies H$. This invariant is maintained: whatever we choose for $\Sigma$, if $\forall \sigma~F \implies H$, $\forall \sigma~F \implies H \land F[\Sigma/\sigma]$. If the algorithm answers ``no solution'' for $H$, because of the invariant, there is no solution for $\forall \sigma~F$. If the solution answers ``solution~$B$'', the ``unsat'' answer for $\textit{SMT}(\neg F[B/b])$ guarantees that $B \models \forall \sigma~F$.

Note the use of two solvers: one SAT for the propositional variables $b$, and one SMT for the state variables $\sigma$ (or $\sigma,\sigma'$). The SAT solver is used incrementally: one only adds new constraints. The SMT solver is always used with the same set of predicates, enabling it to cache theory lemmas.

\subsection{Performance Improvements}
The algorithm in the preceding subsection is sound, complete and terminating. Yet, experiments have shown that it tends to generate useless iterations. 
One reason is that the system may iterate across instances $B$ that yield the same formula $T[B/b]$ up to a permutation of the $C_i$ disjuncts. Another is that the system may generate empty disjuncts $C_i$, or more generally disjuncts that are subsumed by the other disjuncts (and are thus useless). We shall explain how to deal with those issues.


\subsubsection{Removal of Permutations}
\label{part:remove_permutations}
We impose that the disjunction $C_1 \lor \dots \lor C_n$ follows a unique canonical ordering. For this, we impose that the vectors of $m$ Booleans $(B_{1,j})_{1 \leq j \leq m}, \dots, (B_{n,j})_{1 \leq j \leq m}$ are in strict increasing order with respect to the lexicographic ordering $\prec_L$ induced by $\false < \true$. This corresponds to $n-1$ constraints $(b_{i,j})_{1 \leq j \leq m} \prec_L (b_{i+1,j})_{1 \leq j \leq m}$, each of which can be encoded over the propositional variables $(b_{i,j})$ as formula $L_{i,1}$ defined as follows:

\begin{itemize}
\item $L_{i,j_0}$ is a formula whose meaning is that $(b_{i,j})_{j_0 \leq j \leq m} \prec_L (b_{i+1,j})_{j_0 \leq j \leq m}$
\item $L_{i,m+1}$ is \false
\item $L_{i,j_0}$ for $1 \leq j_0 \leq m$ is defined using $L_{i,j_0+1}$ as follows: $(\neg b_{i,j_0} \land b_{i+1,j_0}) \lor ((b_{i,j_0} \Rightarrow b_{i+1,j_0}) \land L_{i,j_0+1})$.
\end{itemize}

All such constraints can be conjoined to the initial value of~$H$.

\subsubsection{Removal of Subsumed Disjuncts}
\label{part:remove_subsumed}
We can replace the SAT-solver used to find solutions for $(b_{i,j})$ by a SMT-solver for theory $\frakT$, in charge of finding solutions for $(b_{i,j})$ and for some auxiliary variables $\sigma_1, \dots, \sigma_n$ (we actually shall not care about the actual values of $\sigma_1, \dots, \sigma_n$). The following constraint expresses that the disjunct $C_{i_0}$ is not subsumed by the disjuncts $(C_i)_{1 \leq i \leq n, i \neq i_0}$:
\begin{equation}
\exists \sigma_{i_0}~ C_{i_0}[\sigma_{i_0}/\sigma] \land \bigwedge_{1 \leq i \leq n, i \neq i_0} \neg C_i[\sigma_i/\sigma]
\end{equation}

It therefore suffices to conjoin to the initial value of $H$ the following constraints, for $1 \leq i_0 \leq n$:
$C_{i_0}[\sigma_{i_0}/\sigma] \land \bigwedge_{1 \leq i \leq n, i \neq i_0} \neg C_i[\sigma_i/\sigma]$.

A variant consists in simply imposing that each of the $C_i$ is satisfiable, thus eliminating useless false disjuncts. For this, one imposes $1 \leq i_0 \leq n$, the constraint $C_{i_0}[\sigma_{i_0}/\sigma]$. Equivalently, one can pre-compute the ``blocking clauses'' over the $b_{i_0,j}$ propositional variable that constrain these variables so that $C_{i_0}$ is satisfiable, and add them as purely propositional constraint. This is the method that we used for the example in Sec.~\ref{sec:step_by_step_example} (we wanted to keep to propositional constraints for the sake of simplicity of exposition).

\subsection{Iterative Refinement of Invariants}
We have so far explained how to compute \emph{any} invariant, with or without imposing a postcondition. If we do not impose a postcondition, the formula {\true}, for instance, can denote a wholly uninteresting invariant; clearly we would like a smaller one. In this section, we shall explain how to obtain \emph{minimal} invariants within the search space.

\subsubsection{For a Fixed Disjunction Size}
\label{sec:iterative_refinement_n}
Let us now assume we have the postcondition~$P$ (if we do not have it, then set $P$ to \true). A natural question is whether one can get a \emph{minimal} inductive invariant of the prescribed form for the inclusion ordering; that is, an invariant $T[B_0/b]$ such that there exists no $B$ such that $T[B/b] \subseteq T[B_0/b]$, by which we denote $\forall \sigma~ T[B/b] \Rightarrow T[B_0/b]$. We shall now describe an iterative algorithm that first obtains any inductive invariant of the prescribed form, and then performs a downwards iteration sequence for the inclusion ordering, until a minimal element is found.

Let us first note that it is in general hopeless to find a global minimum~$B_0$, that is, one such that $\forall B~ T[B_0/b] \subseteq T[B/b]$, for there may exist incomparable minimal elements. For instance, consider the program:

\begin{lstlisting}
float i = 0;
while(random()) {
  i = i+1;
  if (i > 2) i = 0;
}
\end{lstlisting}

The least inductive invariant of this loop, for variable $i$, is the set of floating-point numbers $\{0, 1, 2\}$. Now assume our set of predicates is $\{ i \leq 0, i \geq 0, i \geq 1, i \leq 1, i \leq 2, i \geq 2 \}$, and take $n=2$; we thus look for disjunctions of two intervals. Two minimal incomparable invariants are $(i \geq 0 \land i \leq 1) \lor (i \geq 2 \land i \leq 2)$, that is, $[0, 1] \cup \{ 2 \}$, and $(i \geq 1 \land i \leq 2) \lor (i \leq 0 \land i \geq 0)$, that is, $[1, 2] \cup \{ 0 \}$.

Let us now assume we have already obtained an invariant $T[B'/b]$ and we wish to obtain a better invariant $T[B/b] \subsetneq T[B'/b]$. This last constraint can be written as the conjunction of:
\begin{enumerate}
\item $T[B/b] \subseteq T[B'/b]$, otherwise said $\forall \sigma~ T[B/b] \Rightarrow T[B'/b]$; such a universally quantified constraint can be handled as explained in Sec.~\ref{part:with_postcond}.

\item $\exists \sigma~  T[B'/b] \land \neg T[B/b]$. Again, as explained in Sec.~\ref{part:remove_subsumed}, one can treat such an existentially quantified constraint by using a SMT-solver instead of a SAT-solver and adding to $H$ an extra variable $\sigma$ and the constraint $T[B'/b] \land \neg T[B/b]$. When an invariant $T[B/b]$ is found, the value $\Sigma$ of $\sigma$ is a witness that this invariant is \emph{strictly included} in $T[B'/b]$.
\end{enumerate}

It is possible to compute a downward iteration sequence until a minimal element is reached: compute any initial invariant $B^{(0)}$, then $B^{(1)} \subsetneq B^{(0)}$ etc. until the system fails to provide a new invariant satisfying the constraints; one then takes the last element of the sequence. The termination condition is necessarily reached, for the $(B^{(k)}_{i,j})_{1 \leq i \leq m, 1 \leq j \leq n}$ Boolean matrices can never be twice the same within the sequence (because of the strict descending property). Furthermore, one can stop at any point $B^{(k)}$ within the sequence and get a (possibly non minimal) inductive invariant.

One can replace point 2 above by a weaker strategy, but with the advantage of operating only on propositional formulas. Note that $B^{(k+1)}$ has at least one component higher than $B^{(k)}$ for the standard ordering $\false < \true$ on the Booleans, for if all components are lower or equal, then $B^{(k+1)} \supseteq B^{(k)}$, which is the opposite direction of what we wish. The strategy is to enforce this condition using $\bigvee_{i,j} (b_{i,j} \land \neg b'_{i,j})$. This is what we used in Sec.~\ref{sec:step_by_step_example}.

\subsubsection{For Varying Disjunction Sizes}
The algorithms described above work for a given disjunction size~$n$. The method for preventing subsumed disjuncts of part Sec.~\ref{part:remove_subsumed} imposes that all $n$ disjuncts are truly needed: it is thus possible that no solution should be found for $n=n_0$ while solutions exist for $n=n_0-1$.

We therefore suggest that, once a minimal invariant $I_{n_0}$ is obtained for $n=n_0$ fixed, one looks for an invariant strictly included in $I_{n_0}$ for $n=n_0+1$. One can choose to stop such iterations when no solutions are found for a given $n$, or when a limit on $n$ or a timeout is reached.

\subsection{Extensions}

\paragraph{Prohibition of Overlapping Modes}
Our algorithms produce disjunctions that cover all reachable states, but that do not define partitions: distinct abstract states may be overlapping. This may be somewhat surprising and counterintuitive.

It is possible to impose that disjuncts should be pairwise disjoint. For any $i$ and $j$, one can impose that $C_i$ and $C_j$ are disjoint by the universally quantified formula $\forall \sigma \neg C_i \lor \neg C_j$. We have explained in the preceding sections how to deal with such universally quantified formulas.

\paragraph{Other Template Forms}
\label{sec:other_forms}
We have described our algorithm for templates of the form $C_1 \lor \dots \lor C_m$ where the $C_i$ are conjunctions constructed from the chosen predicates, but the algorithm is not specific to this template shape. Instead of disjunctive normal form, one could choose conjunctive normal form, for instance, or actually any form~\cite{Srivastava:2009:PVU:1543135.1542501}, though reductions of the search space such as those from Sec.~\ref{part:remove_permutations} or \ref{part:remove_subsumed} may be more difficult to define.

\paragraph{Predicate Choice}
Our method is based on predicate abstraction; so far we have not discussed methods for obtaining the predicates, beyond the obvious syntactic detection. In many systems based on predicate abstraction, one uses \emph{counterexample-based abstraction refinement} (CEGAR): from an abstract trace violating the specification, but not corresponding to a concrete trace violating the specification, one derives additional predicates for refining the system. 
Because we did not implement such refinement, we shall only give a rough description of our CEGAR method.

If there is no inductive invariant built from the requested template that can prove the desired postcondition, the algorithm from Sec.~\ref{part:with_postcond} will end up with an unsatisfiable constraint system. This system is unsatisfiable because of the postcondition constraints (otherwise, in the worst case, one would obtain a solution yielding the {\true} formula); relevant postcondition constraints can be obtained from an unsatisfiable core of the constraint system. One can then try removing such constraints one by one until the constraint system becomes satisfiable again. Any solution of this relaxed constraint system defines an inductive invariant, but one that does not satisfy the postcondition. As with the usual CEGAR approach, one could try generating test traces leading from the initial states to the complement of the postcondition and staying within the invariant; if the postcondition holds, such searches are unsuccessful and yield interpolants from which predicates may be mined.

\section{Step-by-step Example of Invariant Inference}
\label{sec:step_by_step_example}
For the sake of simplicity of exposition, in this section we have restricted ourselves to pure propositional constraints on the $b_{i,j}$, and satisfiability modulo the theory of linear integer arithmetic for the combination of the $b_{i,j}$ and the state variables. We consider the following simple program.

\begin{lstlisting}
int b, i=0, a; /* precondition a > 0 */ 
while (i < a) {
  b = random();
  if (b)
    i = i + 1;
}
\end{lstlisting}

The predicates are $\{ \pi_1, \dots, \pi_8\} \defn \{i = 0, i < 0, i > 0, i = a, i < a, i > a, b, \neg b\}$. The state variable $\sigma$ stands for $(i, a, b)$. For the sake of simplicity, we model $i$ and $a$ as integers in $\bbZ$, and $b$ as a Boolean. We assume the loop precondition $S \defn i = 0 \land a \geq 1$. The loop condition is $C \defn i < a$, and the transition relation is $T \defn (b' \land i' = i+1) \lor (\neg b' \land i' = i)$. We choose $n=2$.

We shall now run the algorithm described in Sec.~\ref{part:with_postcond} with the iterative refinement of Sec.~\ref{sec:iterative_refinement_n}. For the sake of simplicity, we shall use none of the improvements described in the preceding sections that need the $H_i$ to contain non propositional variables: no removal of subsumed disjuncts as described in Sec.~\ref{part:remove_subsumed} and no strict inclusion enforcement as described in Sec.~\ref{sec:iterative_refinement_n}.

We initialize $H$ as follows: $H_1$ contains Boolean constraints on $(b_{i,j})_{1 \leq i \leq 2, 1 \leq j \leq 8}$
\begin{itemize}
\item That prevent $C_1$ and $C_2$ from being unsatisfiable, using blocking clauses as explained in Sec.~\ref{part:remove_subsumed}: one cannot have both $i=0$ and $i > 0$, and so on.
\item That force $(b_{1,j})_{1 \leq j \leq 8} \prec_L (b_{2,j})_{1 \leq j \leq 8}$ for the lexicographic ordering $\prec_L$ on Boolean vectors (this avoids getting the same disjunction twice with the disjuncts swapped).
\end{itemize}
Let us now see the constraint solving and minimization steps.

\begin{enumerate}
\item
We perform SAT-solving on $H_1$ and obtain a satisfying assignment
$B^{(1)}_{1, 1} = \true, B^{(1)}_{1, 2} = \false, B^{(1)}_{1, 3} = \false, B^{(1)}_{1, 4} = \true, 
  B^{(1)}_{1, 5} = \false, B^{(1)}_{1, 6} = \false, B^{(1)}_{1, 7} = \true, B^{(1)}_{1, 8} = \false, 
  B^{(1)}_{2, 1} = \true, B^{(1)}_{2, 2} = \false, B^{(1)}_{2, 3} = \false, B^{(1)}_{2, 4} = \true, 
  B^{(1)}_{2, 5} = \false, B^{(1)}_{2, 6} = \false, B^{(1)}_{2, 7} = \false, B^{(1)}_{2, 8} = \true$.
This corresponds to the invariant-candidate $T[B^{(1)}/b]$, that is,
$(i = 0 \land i =a \land b) \lor (i =0 \land i =a \land \neg b)$.

Now is this invariant-candidate truly an inductive invariant? It is not, because it does not contain the whole of the set of initial states. SMT-solving on $S \land \neg T[B^{(1)}/b]$ gives a solution $\Sigma_1 \defn (i = 0, a = 1, b = \false)$. We therefore take $H_2 \defn H_1 \land F[\Sigma_1/\sigma]$.

\item
A satisfying assignment $B^{(2)}$ of~$H_2$ yields the invariant candidate
$(i=0\land i=a\land b)\lor
   (i=0\land i<a\land b)$. Again, SMT-solving shows this is not an invariant because it does not contain the initial state $\Sigma_2 \defn (i = 0, a = -1, b = \false)$.  We therefore take $H_3 \defn H_2 \land F[\Sigma_2/\sigma]$.

\item
A satisfying assignment $B^{(3)}$ of~$H_3$ yields the invariant candidate
$(i=0\land i=a \land b)\lor (i=0\land i<a)$. SMT-solving shows this is not inductive, since it is not stable by the transition $\Sigma_3 \defn (i=0, a=1, b=\false, i'=1, b'=\true)$.  We therefore take $H_4 \defn H_3 \land F[\Sigma_3/\sigma]$.

\item
A satisfying assignment $B^{(4)}$ of~$H_4$ yields the invariant candidate
$(i=0\land i<a\land \neg b)\lor b$. SMT-solving shows this is not inductive, since it is not stable by the transition $\Sigma_4 \defn (i=1, a=3, b=\true, i'=1, b'=\false)$. We therefore take  $H_5 \defn H_4 \land F[\Sigma_4/\sigma]$.

\item
A satisfying assignment $B^{(5)}$ of~$H_5$ yields the invariant candidate
$(i=0\land i<a)\lor (i>0\land i=a\land b)$. SMT-solving shows this is not inductive, since it is not stable by the transition $\Sigma_5 \defn (i=0, a=2, b=\false, i'=1, b'=\false)$. We therefore take $H_6 \defn H_5 \land F[\Sigma_5/\sigma]$.
\item
A satisfying assignment $B^{(6)}$ of~$H_6$ yields the invariant candidate
$I_1 \defn (i=0\land i<a)\lor i>0$. SMT-solving shows this is an inductive invariant, which we retain. We however would like a \emph{minimal} inductive invariant within our search space. As described at the end in Sec.~\ref{sec:iterative_refinement_n}, we take $H_7$ the conjunction of $H_6$ and a propositional formula forcing at least one of the $b_{i,j}$ to be {\true} while $B^{(6)}_{i,j}$ is \false. Furthermore, as described in point~1 of Sec.~\ref{sec:iterative_refinement_n}, we now consider $F_2 \defn F \land (T \Rightarrow I_1)$, which ensures that we shall from now on only consider invariants included in~$I_1$.

\item
A satisfying assignment $B^{(7)}$ of~$H_7$ yields the invariant candidate
$(i>0\land i=a\land b)\lor i<a$. SMT-solving shows this is not included in $I_1$, using $\Sigma_7 \defn (i=-47, a=181, b=\true)$. We therefore take $H_8 \defn H_7 \land F_2[\Sigma_7/\sigma]$.

\item
$H_8$ has no solution. $I_1$ is thus minimal and the algorithm terminates.
\end{enumerate}

A postcondition for this loop is thus $I_1 \land \neg (i < a)$, thus $i>0\land i=a$. Note that the method did not have to know this postcondition in advance in order to prove it.

\section{Construction of the Abstract Automaton}
\label{part:construct_automaton}
We can now assume that the set of reachable states is defined by a formula $I = I_1 \lor \dots \lor I_n$, with each formula $I_i$ meant to define a state $q_i$ of the abstract automaton.

To each couple of states $(q_i,q_j)$ we wish to attach an input-output relation expressed as a formula $\tau_{i,j}$ with variables $\calI$, ranging over the set of possible current values of the inputs and $\calO$ over the set of possible current values of the outputs. 

Recall that $T$ is a formula expressing the transition relation of the reactive node, over variables $\calI$ (inputs), $\sigma$ (preceding state), $\sigma'$ (next state) and $\calO$ (outputs). Then the most precise transition relation is:
\begin{equation}
\tau_{i,j} \defn \exists \sigma,\sigma'~ I_i \land I_j[\sigma'/\sigma] \land T
\end{equation}

Any over-approximation of this relation is a sound transition relation for the abstract automaton. If we have a quantifier elimination procedure for the theory in which $T$ and the $I_i$ are expressed, then we can compute the most precise $\tau_{i,j}$ as a quantifier-free formula; but we can also, if needed, use an approximate quantifier elimination that yields an over-approximation.

\lstset{language=Lustre}
Let us consider, as an example, the following Lustre node. It has a single integer input \lstinline|dir| and a single integer output \lstinline|out|. If \lstinline|dir| is nonzero, then it is copied to \lstinline|out|; else \lstinline|out| decays to zero by one unit per clock cycle:
\begin{lstlisting}
node clicker(dir : int) returns (out : int);
let
	out = if dir >= 1
              then dir
              else if dir <= -1
                   then dir
                   else 0 -> if pre out <= -1
                             then (pre out) + 1
                             else if pre out >= 1
                                  then (pre out) - 1
                                  else 0;
tel.
\end{lstlisting}

In mathematical notation, let us denote \lstinline|dir| by $d$, \lstinline|pre out| by $o$ and \lstinline|out| by $o'$. The state consists in a single variable $o$, thus $\sigma$ is the same as $o$. The transition relation then becomes
\begin{equation}
T \defn \left\{\begin{array}{l}
  (d \neq 0 \land o'=d) \lor (d = 0 \land o \geq 1 \land o' = o-1)\\
                              \lor (d = 0 \land o \leq -1 \land o' = o+1)
                              \lor (d = 0 \land o ' = o = 0)
\end{array}\right.
\end{equation}
Suitable predicates are $\{o \leq -1, o = 0, o \geq 1 \}$, thus defining the set of reachable states as a partition $I_{-1} \lor I_0 \lor I_1$ where $I_{-1} \defn o \leq -1$, $I_0 \defn o = 0$,  $I_1 \defn o \geq 1$.

Let us compute $\tau_{0,1} \defn \exists o,o'~ I_0 \land I_1[o'/o] \land T$, that is, $\exists o,o' o=0 \land o'\geq 1 \land T$: we obtain $d > 0$. More generally, by computing $\tau_{i,j}$ for all $i,j \in \{-1,0,1\}$, we obtain the automaton below; the initializers (left hand side of the Lustre operator $\rightarrow$) define the initial state~$q_0$.

\begin{center}
\begin{tikzpicture}[->,auto, node distance=2.5cm]
\node[state,initial by diamond] (q0) {$q_0$};
\node[state] (qm) [left of=q0] {$q_{-1}$};
\node[state] (qp) [right of=q0] {$q_1$};
\path (qp) edge [loop right] node {$d \geq 0$} (qp);
\path (qm) edge [loop left] node {$d \leq 0$} (qm);
\path (q0) edge node[above] {$d > 0$} (qp);
\path (q0) edge node[above] {$d < 0$} (qm);
\path (qp) edge [bend left] node[below] {$d = 0$} (q0);
\path (qm) edge [bend right] node[below] {$d = 0$} (q0);
\path (qp) edge [bend right=60] node[above] {$d < 0$} (qm);
\path (qm) edge [bend right=60] node[below] {$d > 0$} (qp);
\end{tikzpicture}
\end{center}

Note that the resulting automaton is nondeterministic: in state $q_1$ (respectively, $q_{-1}$), representing $o > 0$ (resp. $o < 0$), if $d = 0$, then one can either remain in the same state or return to the initial state~$q_0$.

\section{Related Work}
There have been many approaches proposed for finding invariants and proving properties on transition systems. \cite{DBLP:conf/asm/Shankar00} surveys earlier ones.

The problem of finding the control structure of reactive nodes written in e.g. Lustre has been studied previously, most notably by B.~Jeannet~\cite{Jeannet_PhD,Jeannet:2003:DPL:820005.820037,DBLP:conf/sas/JeannetHR99}, but with respect to a property to prove: the control structure is gradually refined until the property becomes provable. This supposes that we know the desired property in advance, which is not always the case in a modular setting: the property may pertain to another module, and may not be easy to propagate back to the current module. The \soft{NBac} tool performs such an analysis using convex polyhedra as an abstract domain. More recent methods for refining the control structure of reactive nodes include~\cite{Balakrishnan_et_al_EMSOFT09}.
We have already proposed some modular abstractions for reactive nodes, but these targeted specific filters with no control structure \cite{Monniaux_CAV05} or needed some precomputation of the control structure~\cite{Monniaux_LMCS10}.

The problem of finding disjunctive invariants has been much studied especially in the context of convex numerical domains, such as polyhedra: if the property to prove is not convex, or relies on a non-convex weakest precondition, then \emph{any} analysis inferring convex invariants will fail. A number of methods have been proposed to infer invariants consisting in finite disjunctions of elements of an abstract domain: some distinguish states according to the history of the computation, as in \emph{trace partitioning} \cite{Rival_Mauborgne_TOPLAS07}, some recombine elements according to some affinity heuristics \cite{Sriram_et_al_SAS06,Popeea:2006:IDP:1782734.1782760}, or decompose the transition relation according to some ``convexity witness'' \cite{DBLP:conf/pldi/GulwaniZ10}. Other methods select predicates with which to split the control state~\cite{DBLP:conf/cav/SharmaDDA11}. Some recent methods leverage the power of modern SMT-solvers to impose convex invariants only at a limited subset of program points, and distinguish all execution paths between them, therefore acting as applying a complete trace partitioning between the points in the distinguished subset~\cite{Monniaux_LMCS10,Gawlitza_Monniaux_ESOP11}; the method in the present article also considers a limited subset of program points (e.g. loop heads), but can infer disjunctive invariants at these points too.

Both polyhedral abstraction and predicate abstraction search for an inductive invariant $I$; then, in order to prove that a certain property $P$ always holds, one shows that $I$ is included in $P$. In all static analyzers by abstract interpretation known to the authors, some form of forward analysis is used: the set of initial states influences the invariant $I$ obtained by the system. In contrast, with $k$-induction, as in the \soft{Kind} tool \cite{HagTin-FMCAD-08}
the initial states play a very limited role (essentially, they invalidate $P$ if there exists a trace of $k$ states starting in an initial state such that one of them does not satisfy~$P$). A known weakness of pure $k$-induction is that it may fail to prove a property because it bothers about bad, but unreachable, states. If one has obtained an invariant $I$ by other methods, one can use it to constrain the system and get rid of these bad, unreachable states. Thus, abstraction-based methods and $k$-induction based methods nicely combine.

The algorithms presented in this article can be seen as a form of minimization constrained by a universally quantified formula $\forall \sigma~F$, achieved by maintaining a formula $H$ such that $\forall \sigma~F \Rightarrow H$, $H$ being a conjunction of an increasingly large number of constraints generated from $F$ ``on demand'': a constraint is added only if it is violated by the current candidate solution. This resembles quantifier elimination algorithms we have proposed for linear real arithmetic~\cite{Monniaux_CAV10}; one difference is that the termination argument is simpler: with a finite number $n$ of Booleans as free variables, a new added constraint suppresses at least  one of the $2^n$ models, thus there can be at most $2^n$ iterations; in comparison the termination arguments for arithmetic involve counting projections of polyhedra.

Reductions from invariant inferences to quantifier elimination, or to minimization constrained by a universally quantified formula, have already been proposed for numerical constraints, where the unknowns are numerical quantities, in contrast to the present work where they are Booleans~\cite{Monniaux_LMCS10}.

Reductions from loop invariant inference in predicate abstraction to Boolean constraint solving were introduced in~\cite{Gulwani_et_al_VMCAI09}, but that work assumed a postcondition to prove, as opposed to minimizing the result. The problem we solve is the same as the one from the later work \cite[Sec.~5]{Srivastava:2009:PVU:1543135.1542501}, but instead of concretely enumerating the (potentially exponential) set of paths inside the program (corresponding to all disjuncts in a disjunctive normal form of the transition relation), each path corresponding to one constraint, we lazily enumerate witnesses for such paths. Unfortunately, we do not have an implementation of the algorithm from \cite{Srivastava:2009:PVU:1543135.1542501} at our disposal for performance comparisons.

More generally, a number of approaches for invariant inference based on constraint solving have been proposed in the last years, especially for reducing numerical invariant inference to numerical constraint solving \cite{Gulwani:2008:PAC:1375581.1375616,DBLP:conf/cav/ColonSS03} or mathematical programming \cite{DBLP:journals/entcs/GoubaultRLLM10}. One difference between these constraint approaches and ours, except that our variable are Boolean and theirs are real, is that we use a \emph{lazy} constraint generation scheme: we generate constraints only when a candidate solution violates them, a method long known in mathematical programming when applying \emph{cuts}. We applied a similar technique for quantifier elimination for linear real arithmetic, using lazy conversions to conjunctive normal form~\cite{Monniaux_CAV10}. A recent \emph{max-policy iteration} considers each path through the loop as a constraint, and lazily selects a combination of paths, using SMT-solving to point the next relevant path~\cite{Gawlitza_Monniaux_ESOP11}.

\section{Conclusion}
We have given algorithms for finding loop invariants, or, equivalently, invariants for reactive nodes, given as templates with Boolean parameters. Using disjunctive invariants for reactive nodes, one obtains an abstraction of the reactive node as a finite automaton with transitions labeled with guards over node inputs.

If a system consists of a number of nodes, then some of these nodes may be replaced by their abstract automaton, resulting in a more abstract system whose behaviors include all behaviors of the original system. This new system can in turn be analyzed by the same method. Thus, our method supports modular and compositional analysis.

We provide the \soft{Candle} tool,
built using the \soft{Yices} SMT-solver and the \soft{Mjollnir} quantifier elimination procedure, which computes abstractions of \soft{Lustre} nodes.